\documentclass{elsart}

\usepackage{epsfig}

\begin{document}

\begin{frontmatter}

\title{Basin Hopping with Occasional Jumping}

\author{Masao Iwamatsu\thanksref{label2}}
\thanks[label2]{ E-mail: iwamatsu@ph.ns.musashi-tech.ac.jp, Tel: +81-3-3703-3111 ext.2382, 
Fax: +81-3-5707-2222}
\address{Department of Physics, General Education Center,
Musashi Institute of Technology,
Setagaya-ku, Tokyo 158-8557, Japan }
\author{Yutaka Okabe}
\address{Department of Physics,
Tokyo Metropolitan University,
Hachioji, Tokyo 192-0397, Japan }

\begin{abstract}
Basin-Hopping (BH) or Monte-Carlo Minimization (MCM) 
is so far the most reliable algorithms in chemical physics
to search for the lowest-energy structure of atomic clusters
and macromolecular systems.  BH transforms the complex
energy landscape into a collection of basins, and explores
them by hopping, which is achieved by random 
Monte Carlo moves and acceptance/rejection using the Metropolis
criterion.  In this report, we introduce the
jumping process in addition to the hopping process 
in BH.  Jumping are invoked when the hopping stagnates by 
reaching the local optima, and are achieved using the 
Monte Carlo move at the temperature $T=\infty$ without rejection.
Our Basin-Hopping with Occasional Jumping (BHOJ) algorithm 
is applied to the Lennard-Jones clusters of several
notoriously difficult sizes.  It was
found that the probability of locating the true global optima
using BHOJ is significantly higher than the original BH.   
\end{abstract}

\end{frontmatter}

\section{Introduction }
\label{sec1}
The Monte Carlo method based on the Metropolis algorithm is 
the most successful and influential stochastic algorithm 
of the 20th century and has been used in variety of 
applications not limited to physics~\cite{Beichl}.  The 
method is 
powerful in an exhaustive search of highly multi-dimensional 
phase space, and, hence, has been routinely used to calculate 
the thermal averaging of statistical physics.

Aside from the applications to statistical physics, 
the Metropolis algorithm has been used as a vehicle for
global optimization, that is, a task to 
search for the lowest minimum point in a rugged landscapes in
a high dimension.  In fact, the simulated annealing (SA)~\cite{Kirkpatrick} 
based on the Metropolis
algorithm is the oldest metaheuristics in global 
optimization.  

In global optimization, a good balance between a 
global search (exploration) and a local search 
(exploitation) is necessary.  Since the Metropolis 
algorithm has only the ability to perform a global search, it is usually 
necessary to augment SA using a more traditional local optimization method
to handle realistic problems~\cite{Wille}.  Later, a method which
combines the Metropolis algorithm and the gradient-based local search
algorithm was proposed.   That is the Monte Carlo minimization 
(MCM) of Li and Sherga~\cite{Li}, and the
basin-hopping (BH) of Wales and Doye~\cite{Wales}. 
Wales and Doye~\cite{Wales}, for example, used their
BH algorithm to study the lowest-energy structures of Lennard-Jones 
clusters that consist of up to 110 atoms successfully.  

In this paper, we propose a new variant of the Basin-Hopping(BH) 
algorithm.  In contrast to another
variant of the BH by Leary and Doye~\cite{Leary1,Leary2} where only
the down-hill search is allowed, we 
borrow the concept of extremal optimization~\cite{Boettcher} or thermal 
cycling~\cite{Mobius}, and introduce the process of {\it jumping}
to enhance the search of a rugged energy landscape.

\section{Jumping in Basin Hopping}
\label{sec2}
The Basin Hopping (BH) algorithm (which is also called the
Monte Carlo plus minimization, MCM)
uses hopping due to Monte Carlo random walks to explore the 
phase space, and gradient-based local optimization to locate 
the local minimum or {\it basin} plus Metropolis criterion 
to accept or reject the move.

Since the energy landscape explored by the usual Monte Carlo move
and immediate relaxation to a nearby basin of attractor using the
gradient method in BH looks
like hopping among basin of attractors, the algorithm is 
termed "Basin Hopping".   In order to enhance the exploration,
the Monte Carlo move in BH consists of a simultaneous displacement 
of all particles in the cluster in contrast to
the usual Monte Carlo method in classical statistical 
mechanics~\cite{Frenkel} where usually a randomly chosen single 
particle is displaced. 

\begin{figure}
\begin{center}
\centerline{\psfig{file=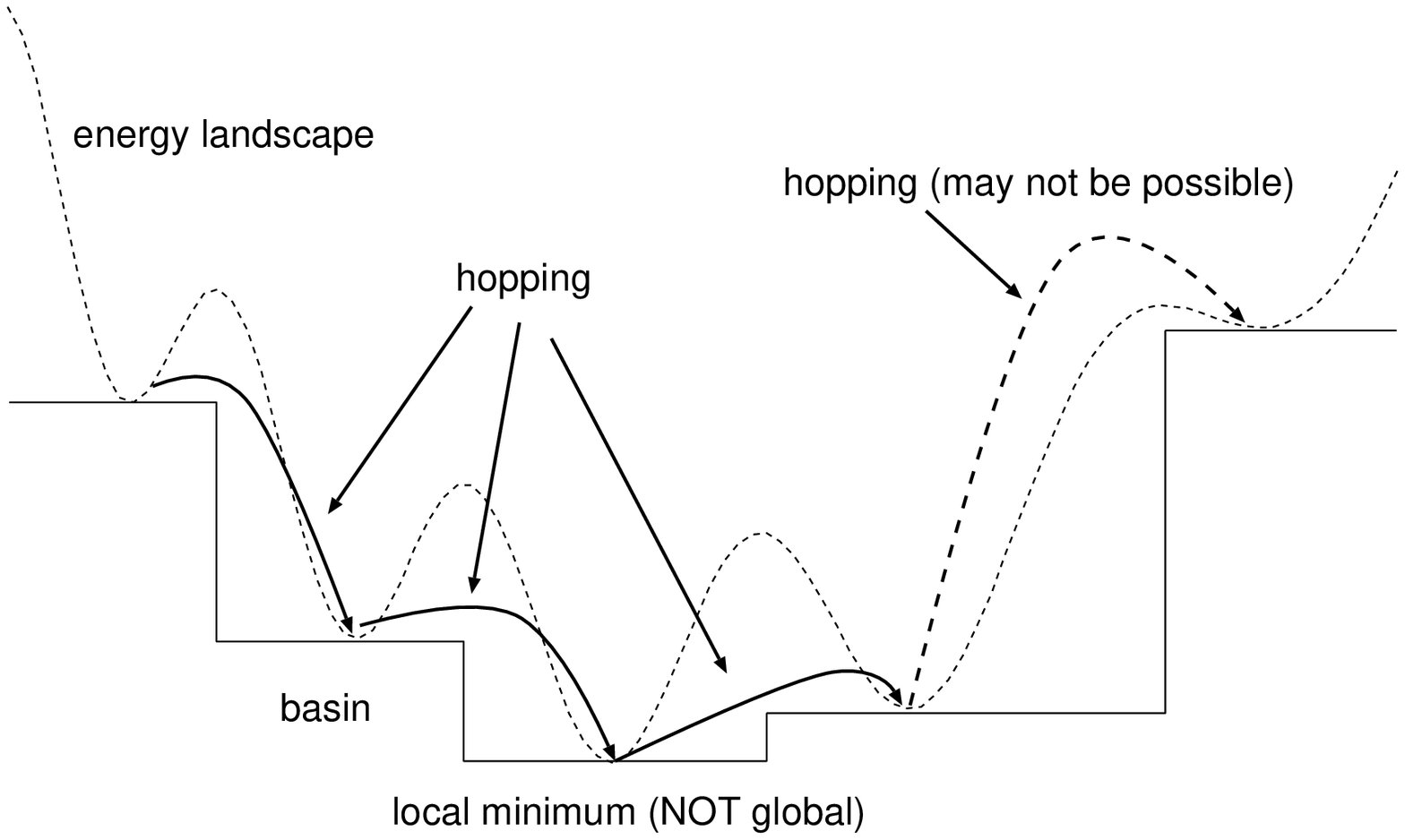,scale=0.5}} 
\caption{
The exploration of the energy landscape using the
BH algorithm. The uphill moves, which uses the 
usual Monte Carlo move may not be possible.}
\end{center}
\label{fig:energy1}
\end{figure}

The BH has been extensively tested~\cite{Wales} for the simplest
benchmark problem for the Lennard Jones clusters 
LJ$_{n}$~\cite{Wille,Hoare,Northby,Deaven,Niesse} where
the total energy of $n$-atom cluster is given by
\begin{equation}
E_{n}=2\sum_{i=1}^{n}\sum_{j=1}^{n}\left(\frac{1}
{r_{ij}^{12}}-\frac{1}{r_{ij}^{6}}\right)
\label{equation}
\end{equation}
where $r_{ij}$ is the distance between two particles $i$ and $j$ within
the cluster.  Even for this simple Lennard-Jones problem, the BH 
clarified the extremely
complex energy landscape for several clusters~\cite{Doye}, and
it even failed to locate the lowest-energy structures for several special
sizes.

For example, the BH
could not locate the lowest minimum of LJ$_{75-77}$, LJ$_{98}$, and
LJ$_{102-104}$ easily~\cite{Wales} when an unbiased search that starts from
a completely random initial configuration of clusters is used. 
In fact, in order to find out
the lowest-energy structure of LJ$_{76}$, Wales and Doye~\cite{Wales}
had to use {\it seeding} to feed the BH the lowest-energy structure 
of smaller LJ$_{75}$ or larger LJ$_{77}$ clusters.  The unbiased
search has an extremely low probability (4\%) of hitting the lowest-energy
structure even when large numbers (100 times) of a fairly 
long run (5000 step) were executed.  The same problem occurred
for LJ$_{102-104}$, and less severely in LJ$_{69}$, LJ$_{78}$ 
and LJ$_{107}$.      

Difficulty occurs when the energy landscape consists of
several large valleys (funnels) instead of one, and 
the funnel corresponding to the
lowest-energy minimum is narrow and separated from other 
funnels by a high barrier~\cite{Doye}.  The Monte Carlo 
trial move plus local minimization and acceptance/rejection 
using the Metropolis criterion is less powerful and time-consuming 
to overcome such a large barrier (Fig. 1).

Recently, a new variant of the basin-hopping algorithm called 
the "Monotonic Sequence Basin-Hopping algorithm" (MSBH) was proposed by 
Leary and Doye~\cite{Leary1,Leary2}.  Their algorithm is essentially the
BH at temperature $T=0$, which allows only downhill moves.  
When the search is stacked at a local minimum, 
the program restarts from a new random configuration.  Therefore, 
MSBH~\cite{Leary1,Leary2}
is essentially the multi-start strategy of a greedy search.  
Naturally, the MSBH algorithm seems less powerful~\cite{Leary2} than 
the original BH algorithm because it is not equipped with 
any mechanism to cross the barrier.

One of the present authors~\cite{Iwamatsu} suggested another 
extension of the BH by introducing the concept of "extremal optimization" 
(EO) proposed by Boettcher and Percus~\cite{Boettcher}.  In this
EO-based basin-hopping (EOBH) algorithm, the Monte Carlo 
move of only one less-bounded particle 
within the cluster is attempted,
and every move is accepted without using the Metropolis criterion.  Therefore
this EOBH is essentially the BH 
at $T=\infty$.  Although the EOBH
can achieve the crossing of high barriers in contrast to 
the MSBH, it
is again less powerful than the original BH because its ability
in terms of a local search is less effective though the method is proven
to be useful to enumerate all the low-energy structures~\cite{Iwamatsu}.

The lesson learned from these two previous exercises is that the 
inclusion of the high-temperature Monte Carlo move at $T=\infty$
will enhance the ability of a global search (exploration), but
the original Metropolis criterion of the Monte Carlo move 
should be retained to maintain the efficiency of a 
local search (exploitation).  Actually,
such a prescription of re-heating or thermal treatment at
high temperatures has been 
repeatedly proposed in the application of the simulated 
annealing (SA)~\cite{Mobius,Ingber}.  For example, 
M\"obius {\it et al}, introduced "Thermal Cycling" and
Ingber~\cite{Ingber} introduced "reannealing" in SA.  In order 
to enhance the ability of crossing the high barrier in the BH, we 
have introduced a re-heating process called "jumping" as 
shown in Fig. 2.

\begin{figure}
\begin{center}
\centerline{\psfig{file=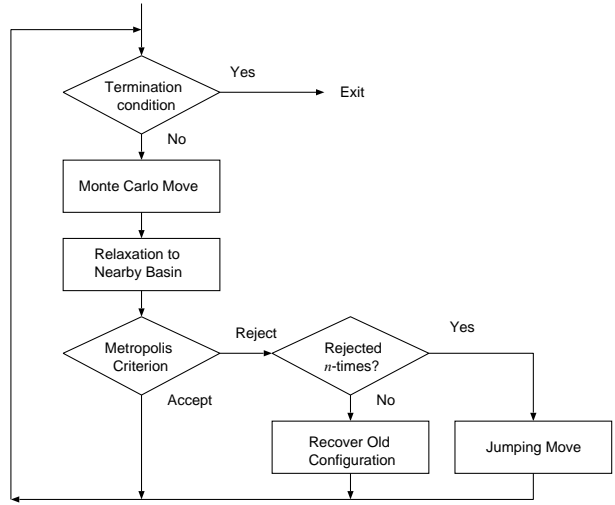,scale=0.7}} 
\caption{
The flow-chart of the basin-hopping with occasional 
jumping (BHOJ) algorithm.
}
\end{center}
\label{fig:flow2}
\end{figure}

Jumping is the Monte Carlo move without relaxation (local 
minimization) at $T=\infty$, and, hence, it is always accepted. 
When the usual Monte Carlo moves are rejected MAX times, the system
is judged to be trapped at the local minimum.  Then the temperature
is raised to $T=\infty$, and the Monte Carlo moves (jumpings) are
executed JMP times
in the search space to allow for the system to escape from the
local minimum (Fig. 3). This move is always 
accepted irrespective of the
increase in the energy because of $T=\infty$
and is called jumping here.
The parameter MAX is used to detect the
entrapment.  The parameter JMP is used to try to climb up the
barrier several times.  If the JMP is too large, the algorithm
is nothing more than simple random multi-start strategy.  It is 
essential to keep the partial memory of the previous 
state~\cite{Mobius}.  So,
the JMP should not be too large.
We will call this version of BH the "Basin-Hopping with
Occasional Jumping" (BHOJ).

\begin{figure}
\begin{center}
\centerline{\psfig{file=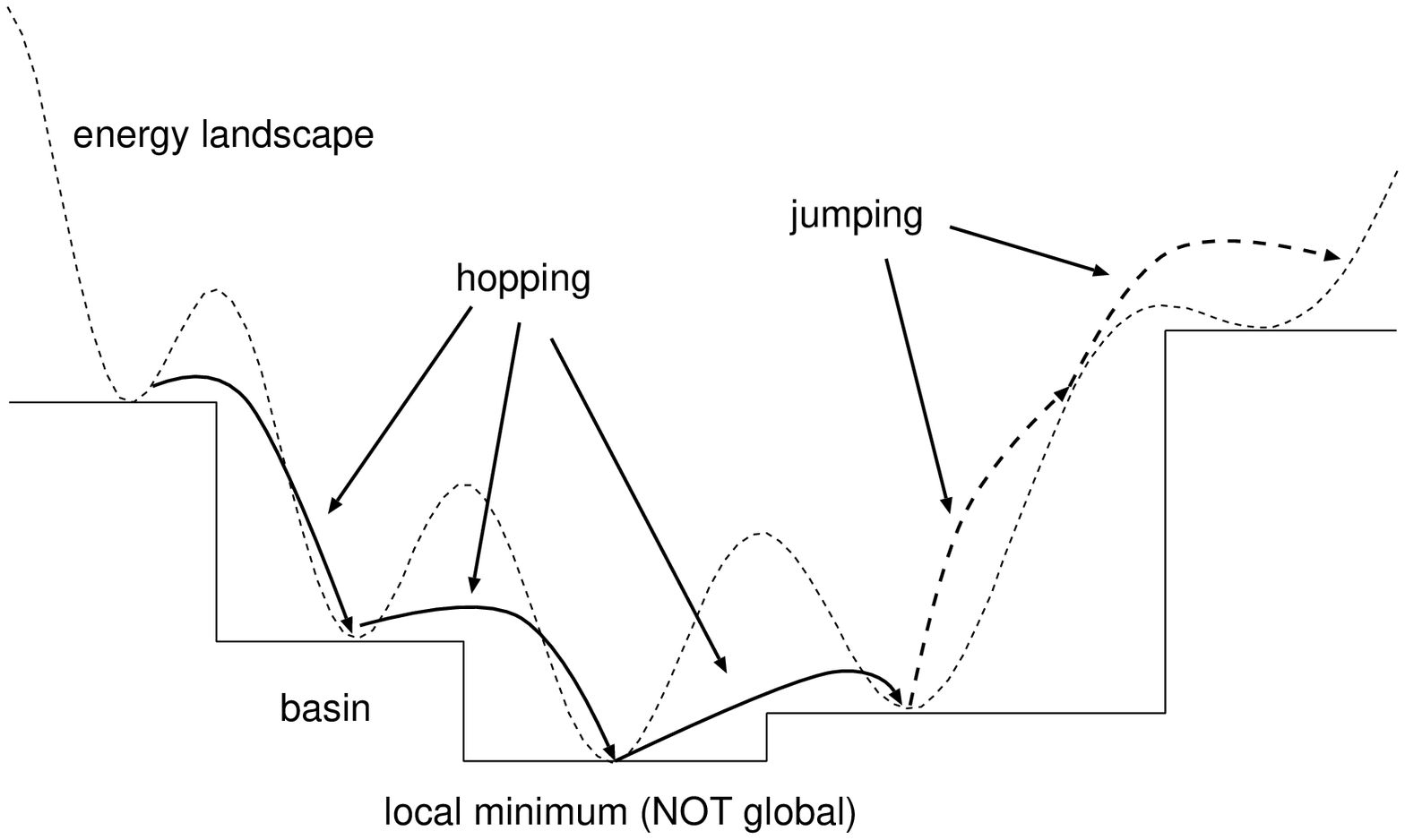,scale=0.5}} 
\caption{
The exploration of the energy landscape (dotted line) using the
BHOJ algorithm. There
are two kinds of uphill moves: one is the hopping (thick solid
arrows) between basins (thin solid line) which uses the 
usual Monte Carlo move followed by the local minimization
and Metropolis acceptance, another is the jumping (thick chain
arrow) using a simple Monte
Carlo move without minimization and rejection.
}
\end{center}
\label{fig:energy2}
\end{figure}

Now, the exploration of phase space in our BHOJ proceeds as
follows (Fig. 3): the rugged energy landscape is transformed into
successive steps of the basin by local minimization.  The downhill
move is simply the descending of the stairs by hopping.  There are,
however, two kinds of uphill moves. One is by climbing the
stairs by hopping using the Metropolis criterion, which is costly 
and may not be effective to climb up the high barrier because
the moves with a large energy difference are rejected by
the Metropolis criterion.  Another move is jumping,
which does not use local minimization and is always accepted, so
the uphill moves by jumping do not use stairs and 
are simply along the surface of the hill
(Fig. 3).  This jumping must be an efficient way to
escape from a local minimum (valley) and to explore the next 
basin of the valley when it is separated by high barriers.

\section{Experiments}
In order to test the performance of our modification of the 
BH with occasional jumping (BHOJ), we have calculated 
the lowest-energy structure of the Lennard-Jones
clusters of a particular size, LJ$_{38}$, LJ$_{75-77}$, LJ$_{98}$, 
and LJ$_{102-104}$ for which the original BH is not effective.  We
conducted a 100 independent unbiased search starting from 100 random
initial structures with the maximum of
5000 iterations which do not include the jumping process.  We also
performed the same experiment using the original BH~\cite{Wales2} with the same
random initial structures.  
The temperature is fixed at $T=0.8$.  Two additional parameters 
MAX and JMP for the BHOJ are arbitrarily
fixed to MAX=10 and JMP=3 or 5 or 7.

Table 1 gives the success rates of 100 unbiased searches.  Our BHOJ
could successfully reproduce the lowest-energy structures in
the literature~\cite{Wales,Cambridge}.  It is apparent that
our BHOJ performs in general better than the original BH.  
The success rate increased twice to five times from the original 
BH to the BHOJ.  
Thus, the ability of exploration is in fact enhanced by the introduction 
of jumping processes.  For the
sake of comparison, we also showed the results of MSBH of 
Leary~\cite{Leary2} and the parallel fast annealing (PFA) 
of Cai {\it et al.}~\cite{Cai2}.

\begin{table}[]
\begin{center}
\caption{
Success rates of original BH~\cite{Wales}, MSBH~\cite{Leary2}, 
PFA~\cite{Cai2} and our BHOJ for selected 
Lennard-Jones clusters which
are notoriously difficult cases to find out the lowest-energy structure.  In
MSBH Leary conducted experiments 1000 times while we conducted experiments
100 times for original BH and BHOJ.
}
\begin{tabular}{crrrrr}
\hline
Cluster & Energy & BH & MSBH & PFA & BHOJ (JMP) \\
\hline
LJ$_{38}$ & -173.928427 & 87/100 & 124/1000 & 39/100 & 96/100 (7) \\
LJ$_{75}$ & -397.492331 & 1/100  & 4/1000   & 2/200  & 5/100 (3) \\
LJ$_{76}$ & -402.894866 & 5/100  & 8/1000   & 2/50   & 10/100 (5) \\
LJ$_{77}$ & -409.083517 & 6/100  &          & 1/50   & 5/100  (7) \\
LJ$_{98}$ & -543.665361 & 10/100 & 6/1000   & 4/100  & 10/100 (3) \\
LJ$_{102}$ & -569.363652 &3/100 & 31/1000  & 9/100   & 16/100 (3) \\
LJ$_{103}$ & -575.766131 & 3/100 &          & 3/30   & 13/100 (5) \\
LJ$_{104}$ & -582.086642 & 3/100 &          & 2/30   & 12/100 (3)\\
\hline
\end{tabular}
\end{center}
\end{table}

\begin{figure}[h]
\centerline{\psfig{file=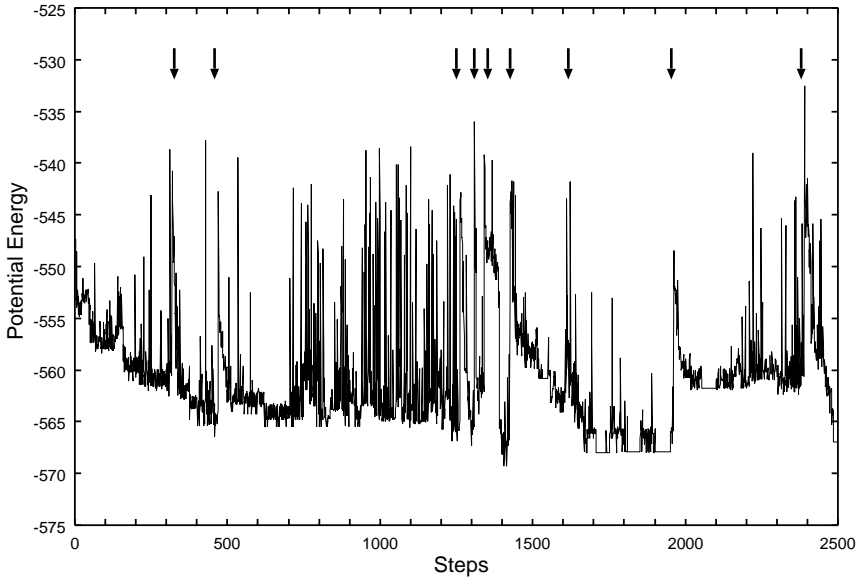,scale=0.6}} 
\caption{
An example of the trajectory of the successive run of BHOJ 
for LJ$_{102}$.  Arrows indicate the positions where the
jumping occurs.
}
\label{fig:trajectory}
\end{figure}

Figure~\ref{fig:trajectory} shows the trajectory of a successful 
run for LJ$_{102}$ when
using the BHOJ.  Jumping occurs only 9 times during the run of
0 to 2500 steps.  However, this jumping may induce
the uphill moves which assist in the exploration of the next 
valley separated from the previous valley by a high barrier, and
our BHOJ can successfully find out the lowest-energy structure
at approximately 1400 steps.

In comparison to the various sophisticated 
methods~\cite{Wolf,Pillardy,Hartke,Rata,Lee}, our basin-hopping
with occasional jumping (BHOJ) is
intuitively appealing and simple to implement.  The performance
of the algorithm seems better than most of the above algorithms.

Finally, in order to further test the performance of our BHOJ, we
have used the BHOJ to search for the lowest-energy
structure of larger clusters even larger than 
LJ$_{148}$~\cite{Romero,Houston}.

\begin{table}[h]
\begin{center}
\caption{
Success rates of original BH~\cite{Wales} and 
our BHOJ for selected larger Lennard-Jones clusters.
}
\begin{tabular}{crrr}
\hline
Cluster & Energy & BH & BHOJ (JMP) \\
\hline
LJ$_{107}$ & -602.007110 & 12/100 & 19/100 (7) \\
LJ$_{185}$ & -1125.493794  & 0/500 &  1/500 (5) \\
LJ$_{186}$ & -1132.669966 & 1/200  & 2/200 (3) \\
LJ$_{187}$ & -1139.455696 & 0/200 & 2/200  (3) \\
\hline
\end{tabular}
\end{center}
\end{table}

Table 2 shows the results for LJ$_{107}$ and LJ$_{185-187}$.  For
LJ$_{107}$ we could confirm the new lowest-energy structure found
by Wales and Doye~\cite{Wales} using our BHOJ.  We could also 
successfully confirm 
the lowest energy -1125.493794 cited in \cite{Houston} found 
by Leary for LJ$_{185}$
which is lower than the previous record -1125.304876 found 
by Hartke~\cite{Hartke,Houston}.
The new lowest-energies -1132.669966 for LJ$_{186}$ and
-1139.455696 for LJ$_{187}$ found by Hartke~\cite{Hartke,Houston}
were also successfully located by our BHOJ though the success rates
of these three cases were very low.

\section{Conclusion}
In this paper we proposed a way to improve the performance
of the basin hopping (BH) algorithm by introducing the jumping
in addition to the hopping.  We call this new algorithm as the
basin-hopping with occasional jumping (BHOJ).  
The jumping is a process of heating
the system and raising the temperature to infinitely 
high which is attempted when the trajectory
in the phase space is trapped at a local minimum.  By jumping, the 
trajectory can climb up high barriers and can explore
the next valley.  Thus the exploratory ability  
of the algorithm is enhanced.

Experiments on benchmark problem of the Lennard-Jones clusters, in particular,
for notorious difficult sizes of 75 to 77 particles 
LJ$_{75-77}$, of 98 particles LJ$_{98}$,
and of 102 to 104 particles LJ$_{102-104}$ reveal
that the proposed BHOJ is really superior to the
original BH.  

This jumping is easy to implement, and 
consumes very little CPU resources.  Any adaptive or
scheduled jumping could be easily incorporated.  The BHOJ
with jumping will be helpful to search for the lowest-energy
structures of larger clusters and more complex clusters with
many body forces.

\section*{Acknowledgments}
One of the author (M.I.) would like to thank Department of Physics, Tokyo
Metropolitan University for providing him a visiting fellowship.


\begin{thebibliography}{22}
\bibitem{Beichl} I. Beichl and F. Sullivan, Computing in Science and Engineering {\bf 2} (2000) 65.
\bibitem{Kirkpatrick} S. Kirkpatrick, C.D. Gelatt, M.P. Vecchi, Science {\bf 220} (1983) 671.
\bibitem{Wille} L.T. Wille, Chem. Phys. Lett. {\bf 133} (1987) 405.
\bibitem{Li} Z. Li, H.A. Scheraga, Proc. Natl. Acad. Sci. U.S.A. {\bf 84} (1987) 6611.
\bibitem{Wales} D.J. Wales, J.P.K. Doye, J. Phys. Chem. A {\bf 101} (1997) 5111.
\bibitem{Leary1} R.H. Leary and J.P.K. Doye, Phys. Rev. E {\bf 60} (1999) R6320.
\bibitem{Leary2} R.H. Leary, J. Global. Opt. {\bf 18} (2000) 367.
\bibitem{Boettcher} P. Boettcher and A. Percus, Artificial Intelligence {\bf 119} (2000) 275.
\bibitem{Mobius} A.M\"obius, A. Neklioudov, A. D\'iaz-S\'anchez, K.H. Hoffmann, A. Fachat, M. Schreiber, Phys. Rev. Lett. {\bf 79} (1997) 4297.
\bibitem{Frenkel} D. Frenkel, B. Smit, {\it Understanding Molecular Simulation
From Algorithm to Application}, Academic Press, 1986.
\bibitem{Hoare} M.R. Hoare, P. Pal, Adv. Phys. {\bf 20} (1971) 161.
\bibitem{Northby} J.A. Northby, J. Chem. Phys. {\bf 87} (1987) 6166.
\bibitem{Deaven} D.M. Deaven, N. Tit, J.R. Morris, K.M. Ho, Chem. Phys. Lett. {\bf 256} (1996) 195.
\bibitem{Niesse} J.A. Niesse, H.R. Mayne, J. Chem. Phys. {\bf 105} (1996) 8428.
\bibitem{Doye} J.P.K. Doye, M.A. Miller, D.J. Wales, J. Chem. Phys. {\bf 111} (1999) 8417.
\bibitem{Iwamatsu} M. Iwamatsu, Proc. 2002 IEEE World Congr. Comput. Intelligence. pp.1180, IEEE Press, 2002.
\bibitem{Ingber} L. Ingber, Mathl. Comput. Modelling {\bf 12} (1989) 967.
\bibitem{Wales2} D.J. Wales, program "gmin", http://www-wales.ch.cam.ac.uk /software.html.
\bibitem{Cambridge} D.J. Wales, "Cambridge Cluster Database", \\http://www-doye.ch.cam.ac.uk/jon/structures/LJ /tables.150.html).
\bibitem{Cai2} W. Cai, H. Jiang, X. Shao, J. Chem. Inf. Comput. Sci. {\bf 42} (2002) 1099.
\bibitem{Wolf} M.D. Wolf and U. Landmann, J. Phys. Chem. A {\bf 102} (1998) 6129.
\bibitem{Pillardy} J. Pillardy, A. Liwo, H.A. Scheraga, J. Phys. Chem. A {\bf 103} (1999) 9370.
\bibitem{Hartke} G. Hartke, J. Comput. Chem. {\bf 20} (1999) 1752.
\bibitem{Rata} I. Rata, A.A. Shvartsburg, M. Horoi, T. Frauenheim, K.W.M. Siu, K.A. Jackson, Phys. Rev. Lett. {\bf 85} (2000) 546.
\bibitem{Lee} J. Lee, I-H. Lee, J. Lee, Phys. Rev. Lett. {\bf 91} (2003) 080201.
\bibitem{Romero} D. Romero, C. Barr\'on, S. G\'omez, Comp. Phys. Commun. {\bf 123} (1999) 87.
\bibitem{Houston} C. Barr\'on, http://www.vcl.uh.edu/$\sim$cbarron/LJ\_ cluster /LJpottable.html (currently down).
\end{thebibliography}
\end{document}